\documentclass[prb,twocolumn,superscriptaddress,showpacs]{revtex4}

\usepackage{graphicx}
\usepackage{dcolumn}
\usepackage{bm}
\usepackage{color}
\begin{document}
\title{Optical study of phonons and electronic excitations in tetragonal Sr$_2$VO$_4$}
\author{J. Teyssier}
\affiliation{D\'epartement de Physique de la Mati\`ere Condens\'ee, Universit\'e de Gen\`eve, Quai Ernest-Ansermet 24, 1211 Gen\`eve 4, Switzerland}
\author{R. Viennois}
\affiliation{D\'epartement de Physique de la Mati\`ere Condens\'ee, Universit\'e de Gen\`eve, Quai Ernest-Ansermet 24, 1211 Gen\`eve 4, Switzerland}
\author{E. Giannini}
\affiliation{D\'epartement de Physique de la Mati\`ere Condens\'ee, Universit\'e de Gen\`eve, Quai Ernest-Ansermet 24, 1211 Gen\`eve 4, Switzerland}
\author{R.~M.~Eremina}
\affiliation{E. K. Zavoisky Physical Technical Institute, 420029
Kazan, Russia}
\author{A. G\"{u}nther}
\affiliation{Experimentalphysik V, Center for Electronic
Correlations and Magnetism, Institute for Physics, Augsburg
University, D-86135 Augsburg, Germany}
\author{J. Deisenhofer}
\affiliation{Experimentalphysik V, Center for Electronic
Correlations and Magnetism, Institute for Physics, Augsburg
University, D-86135 Augsburg, Germany}
\author{M.~V.~Eremin}
\affiliation{Kazan (Volga region) Federal University, 420008 Kazan,
Russia}
\author{D. van der Marel}
\affiliation{D\'epartement de Physique de la Mati\`ere Condens\'ee, Universit\'e de Gen\`eve, Quai Ernest-Ansermet 24, 1211 Gen\`eve 4, Switzerland}
\date{\today}
\begin{abstract}
We report on the optical excitation spectra in Sr$_2$VO$_4$. The phonon modes are assigned and their evolution with temperature is discussed in the frame of the different phase transitions crossed upon cooling. Besides the expected infrared-active phonons we observe two additional excitations at about $290$~cm$^{-1}$ and $840$~cm$^{-1}$ which could correspond to electronic transitions of the V$^{4+}$ ions. Our experimental results are discussed in the context of recent experimental and theoretical studies of this material with a unique spin-orbital ground state.
\end{abstract}
\maketitle
\section{Introduction}

Sr$_2$VO$_4$ has the same layered crystal structure as the parent compound of the high T$_c$ superconductors La$_2$CuO$_4$ \cite{Cyrot1990,Rey1990} and quasi-two-dimensional electronic behaviors. Both materials are Mott-Hubbard insulators with one electron on the Vanadium-site in the case of Sr$_2$VO$_4$, and one hole per copper-site for La$_2$CuO$_4$. Based on these similarities several theoretical groups have suggested that doped Sr$_2$VO$_4$ could be superconducting \cite{Pickett1989,Singh1991,Arita2007}. The excruciating difficulty to synthesize a polycrystalline powder of the pristine material gives less flexibility than in the cuprates to play with the stochiometry. While superconductivity has been elusive, Sr$_2$VO$_4$ passes as a function of temperature through a number of different electronic phases which are not fully understood and which have not been observed in the cuprate family. Above 127 K the material is paramagnetic. Below  97 K the system enters a different magnetic and orbital state, which is not ferromagnetic and the nature of which is the main subject of the present paper. A phase-coexistence is found between 98 K and 127 K \cite{Zhou2007}.\\

The most important difference in electronic structure of La$_2$CuO$_4$ and Sr$_2$VO$_4$ results from the
different ground state degeneracy of open 3d shell in both materials: The hole on the Cu-atom in La$_2$CuO$_4$ occupies a $3d_{x^2-y^2}$ orbital. The single electron on the V-ion in Sr$_2$VO$_4$ occupies the degenerate set of $3d_{xz}$ and $3d_{yz}$ orbitals. This orbital degree of freedom profoundly affects the physical properties, as we will see below.
\\
Based on magnetic susceptibility data it was believed that the system undergoes an antiferromagnetic (AFM)
transition with a sample-dependent N\'eel temperature ranging between
10~K and 100~K, although no signs of long-range order could be found
by neutron scattering experiments \cite{Cyrot1990}.
Subsequent works reported a N\'{e}el temperature of
40K \cite{Nozaki1991} and a strong dependence of the occurrence of
these anomalies on the stoichiometry of the
samples \cite{Suzuki1992}. Only recently, it was shown that the system exhibits a phase
transition at $T_1\simeq 100$~K upon cooling, where the ratio of the
tetragonal lattice parameters $c/a$ increases abruptly while the
magnetic susceptibility drops concomitantly \cite{Zhou2007}. Hence,
the phase below $T_1$ was interpreted as antiferromagnetically and orbitally ordered \cite{Zhou2007}.
Theoretical scenarios predict a stripe-like orbital and collinear
AFM spin ordering \cite{Imai2005}, strong competition between ferromagnetism and antiferromagnetism \cite{Imai2005}, or a magnetically hidden order of
Kramers doublets due to the formation of \textit{magnetic octupoles}
mediated by spin-orbit coupling \cite{Jackeli2009}. Recent inelastic neutron
scattering studies revealed a splitting of the highest lying doublet
of the V$^{4+}$ ions persisting up to 400~K \cite{Zhou2010}, which indicates the presence of a finite dipolar magnetic moment and is difficult to reconcile with a purely octupolar magnetic order. We recently suggested that the system can be described in terms if an alternating spin-orbital order below the N\'{e}el temperature and derived the corresponding energy levels scheme for the V$^{4+}$ ions \cite{Eremin2011}.
\\

Here we investigate the low energy optical excitation spectrum in tetragonal Sr$_2$VO$_4$. The phonon modes have been identified in the optical spectrum by comparison with the isomorphic compounds Sr$_2$TiO$_4$ \cite{Burns1988,Fennie2003} and La$_2$NiO$_4$ \cite{Pintschovius1989}. Their evolution upon cooling, crossing the different structural and (or) magnetic transitions, supports the scenario of a long range orbital ordering. In addition to optical phonon bands we
could identify two excitations. The first one is in the phonon energy range (290 $cm^{-1}$) but does not correspond to any identified phonon. It shows a very strong temperature dependence at the temperature $T_1$ mentioned above. The second one (840 $cm^{-1}$) is too high to be a phonon and could correspond to the above mentioned excitation observed by neutron scattering \cite{Zhou2010}. These two excitations correspond to excitations energies proposed in Ref.\onlinecite{Eremin2011} as a result of the alternating spin-orbital ordered ground state.\\

\section{Sample preparation and experimental details}

Ceramic samples were synthesized from reduction  of homemade Sr$_4$V$_2$O$_9$ precursor using Zr as reducing agent \cite{Viennois2010}. The reaction was done in a quartz tube under vacuum at 950~$^o$C. The procedure was repeated several times with intermediate grinding. With this method, almost single-phase
tetragonal Sr$_2$VO$_4$ can be obtained. Lattice parameters were extracted from Rietveld refinements of the powder diffraction pattern (a=3.8349 {\AA}, c=12.5646 {\AA}) and traces of the orthorhombic phase of Sr$_2$VO$_4$ as well as Sr$_3$V$_2$O$_7$ were found within 3\%. \\

\begin{figure}
  \includegraphics[width=8.5cm]{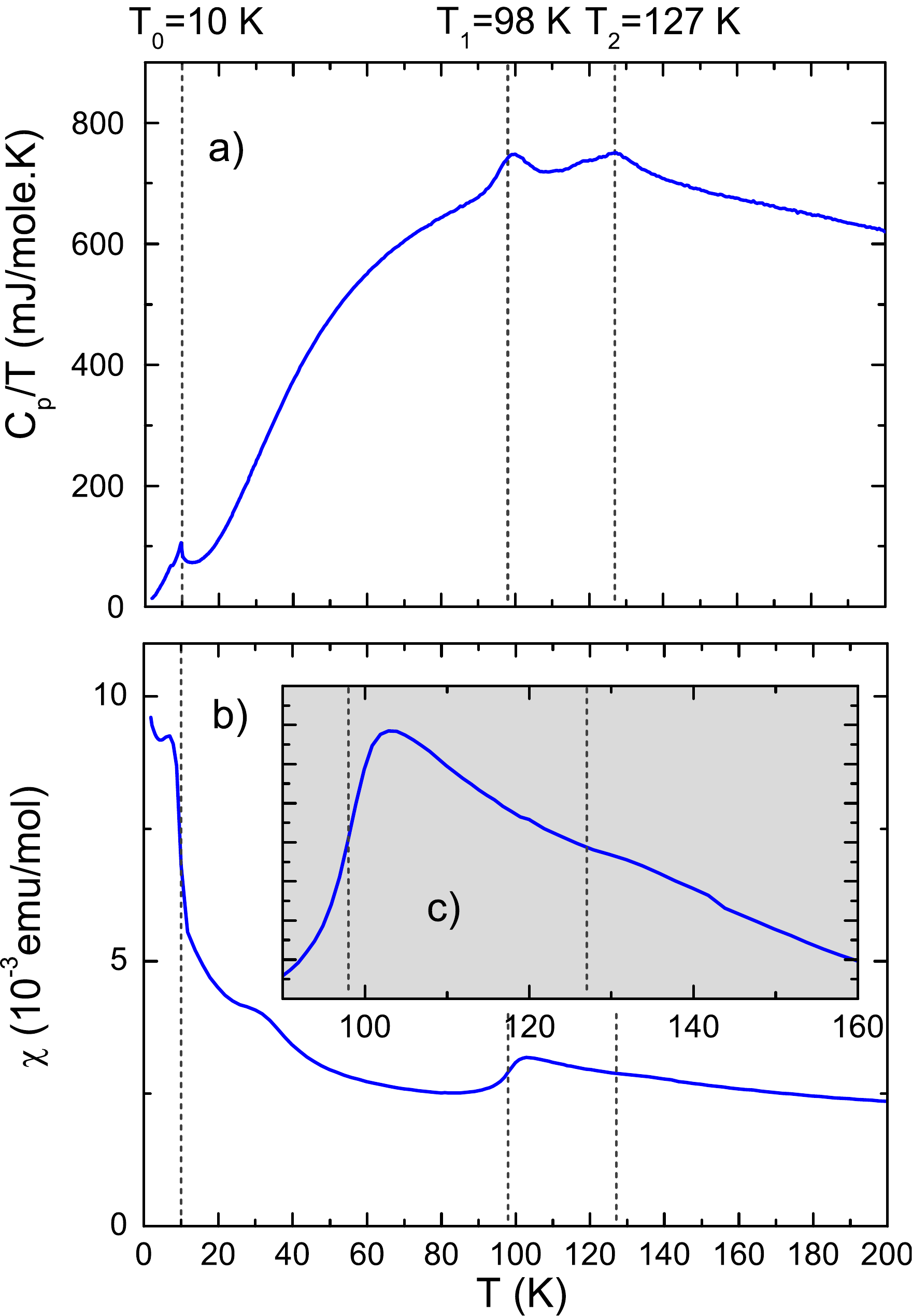}\\
  \caption{a) Specific heat divided by temperature, b) and c) Magnetic susceptibility measured in $0.2$ T.}\label{phys}
\end{figure}

The specific heat exhibits three peaks upon cooling (Fig.~\ref{phys}a). A peak in the specific heat associated to a weak anomaly in the susceptibility at $T_2=127$~K (Fig.~\ref{phys} b) coincides with the reported onset of orbital order and a coexistence regime of the low- and high-temperature tetragonal phase for $T_1\leq T\leq T_2$ \cite{Zhou2007}. The sharp drop at $T_1=98$~K reveals the transition to the low temperature tetragonal phase. A peak at $10$~K indicates the transition to a phase with a weak ferromagnetic moment. \textcite{Nozaki1991} have measured
a small hysteresis loop at $5$~K with a ferromagnetic moment of about $10^{-4}$ $\mu_B$ and a non-vanishing magnetic moment is
also predicted by several theoretical studies \cite{Pickett1989,Imai2005,Arita2007b}. A further broad anomaly, also visible at 35~K in the magnetic
susceptibility, was reported earlier \cite{Nozaki1991}. Since this feature varies rather strongly from one sample to another, it may not be an intrinsic property of tetragonal Sr$_2$VO$_4$. The peaks present in the magnetic susceptibility around 35~K could be attributed to Sr$_3$V$_2$O$_7$ as a secondary phase \cite{Fukushima1993}. The Curie like behavior of the magnetic susceptibility below the N\'{e}el temperature is very likely dominated by impurities \cite{Eremin2011}.\\

The reflectivity of the sample was measured in the infrared spectral range ($12$~meV and $0.8$~eV) using a Fourier transform spectrometer and the complex dielectric function was measured in the range from 0.8-4.5 eV using spectroscopic ellipsometry. For both measurements, the sample was mounted in a helium flow cryostat allowing measurements from room temperature down to 13 K. Absolute reflectivity was obtained by calibrating the signal against an \emph{in situ} evaporated gold film on the sample surface.

\section{Experimental results and discussion}

\begin{figure*}
  \includegraphics[width=14cm]{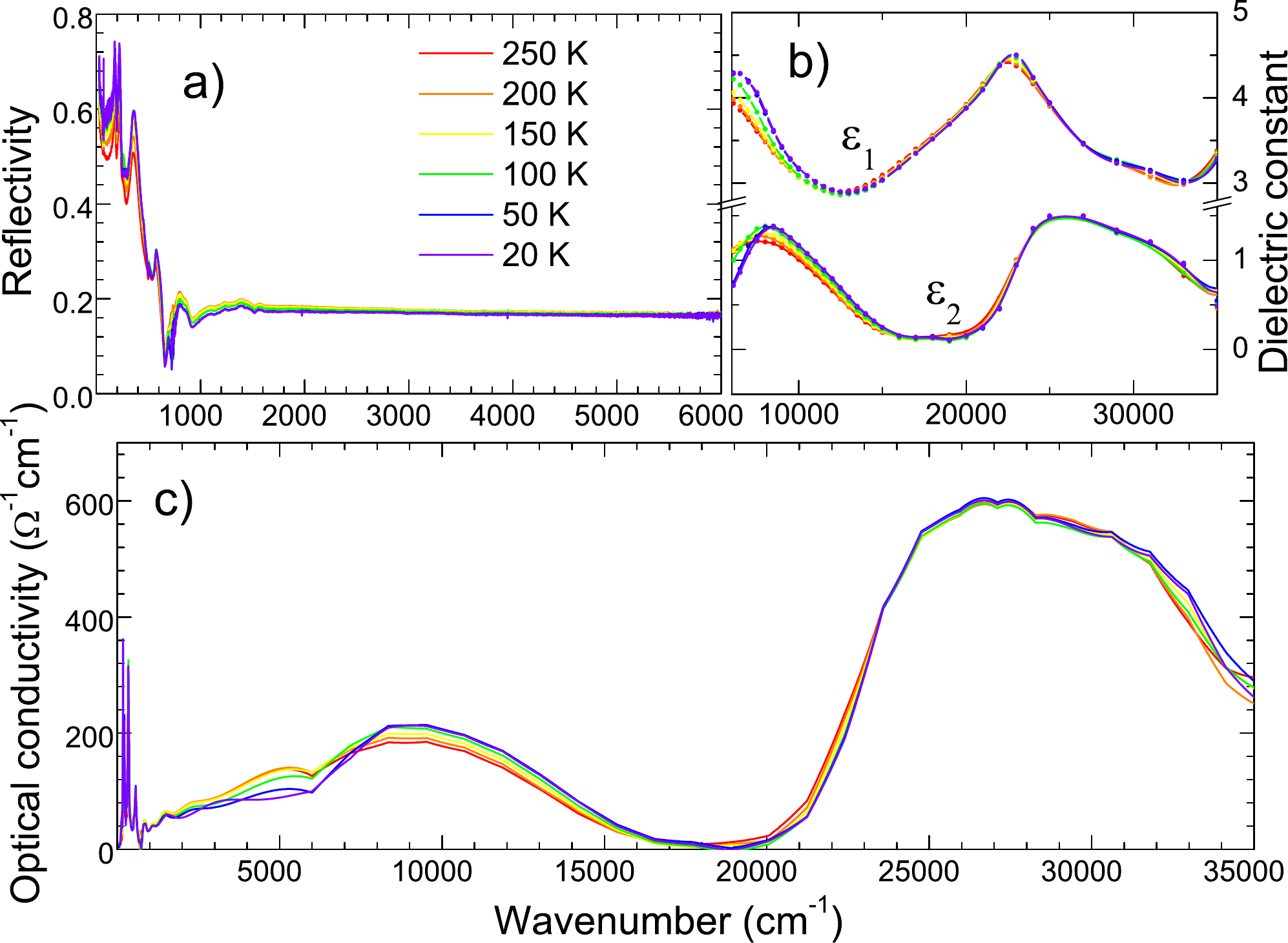}\\
  \caption{a) Reflectivity measured in far- and near-infrared, b) Real ($\epsilon_1$) and imaginary ($\epsilon_2$) part of the dielectric function measured by ellipsometry, c) Real part of the optical conductivity ($\sigma_1$) for selected temperatures.}\label{optics}
\end{figure*}

\begin{figure}
  \includegraphics[width=8.5cm]{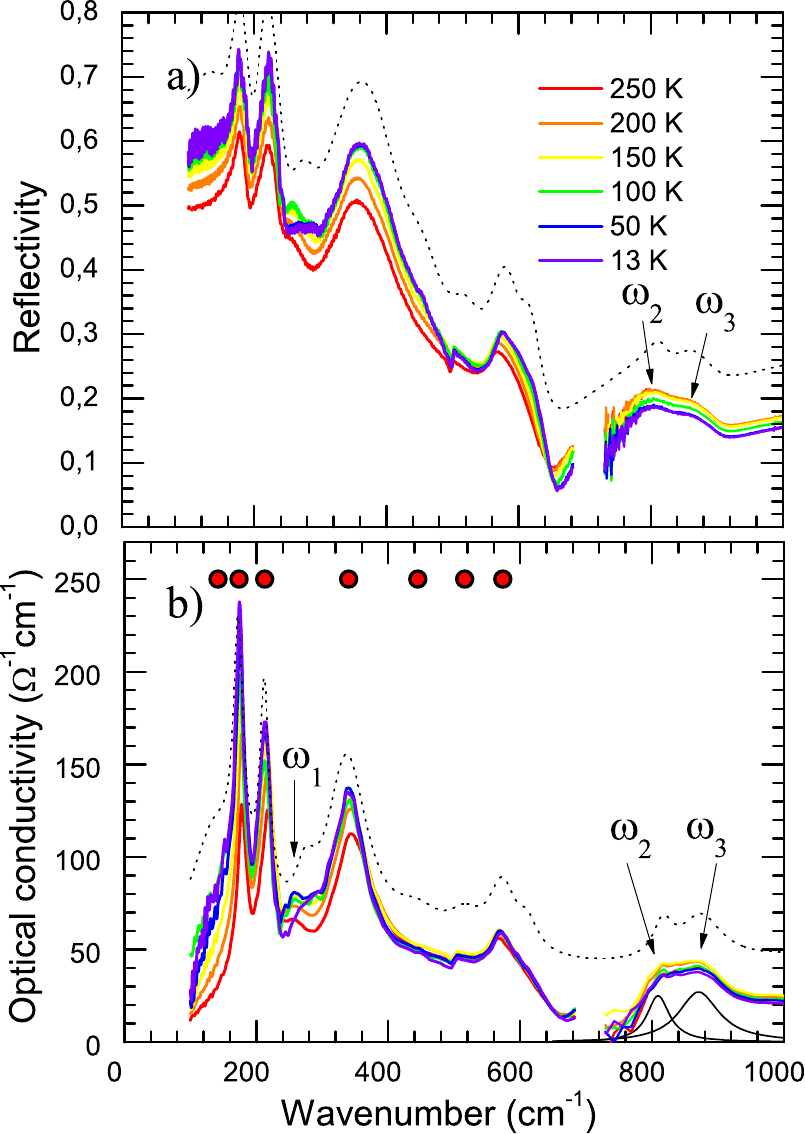}\\
  \caption{Low energy part of the reflectivity (panel a) and optical conductivity (panel b) for different temperatures. The optical phonon frequencies are indicated by filled circles. Inter-orbital optical excitations $\omega_1$, $\omega_2$ and $\omega_3$ are marked by arrows. The dotted line corresponds to the Lorentz fit at 13K (dotted curves were shifted for reasons of clarity)}\label{fir}
\end{figure}

The experimental reflectivity is plotted in Fig. \ref{optics} a). The reflectivity, also measured in the far infrared in a magnetic field as high as 7 T, did not reveal any field dependence of the main features of the spectrum. The real and imaginary parts of $\epsilon(\omega)$ obtained using ellipsometry at selected temperatures are shown in Fig.~\ref{optics}b). In order to obtain the optical conductivity over the full spectral range (Fig.~\ref{optics}c), we used a variational routine \cite{Kuzmenko2005} yielding the Kramers-Kronig consistent dielectric function that reproduces all the fine details of the infrared reflectivity data while {\em simultaneously} fitting to the complex dielectric function in the visible and UV-range. This procedure anchors the phase of the infrared reflectivity to the phase at high energies. The temperature evolution of the optical conductivity in the visible range is in good agreement with previous measurements on epitaxial thin films \cite{Matsuno2005}. Fig.~\ref{fir} shows the low energy part of the reflectivity (panel a) and the corresponding optical conductivity (panel b). The phonon modes that have been identified (see section \ref{phonons}) are marked by solid dots.

\subsection{Phonons}\label{phonons}

\begin{figure}[htp]
  \includegraphics[width=9cm]{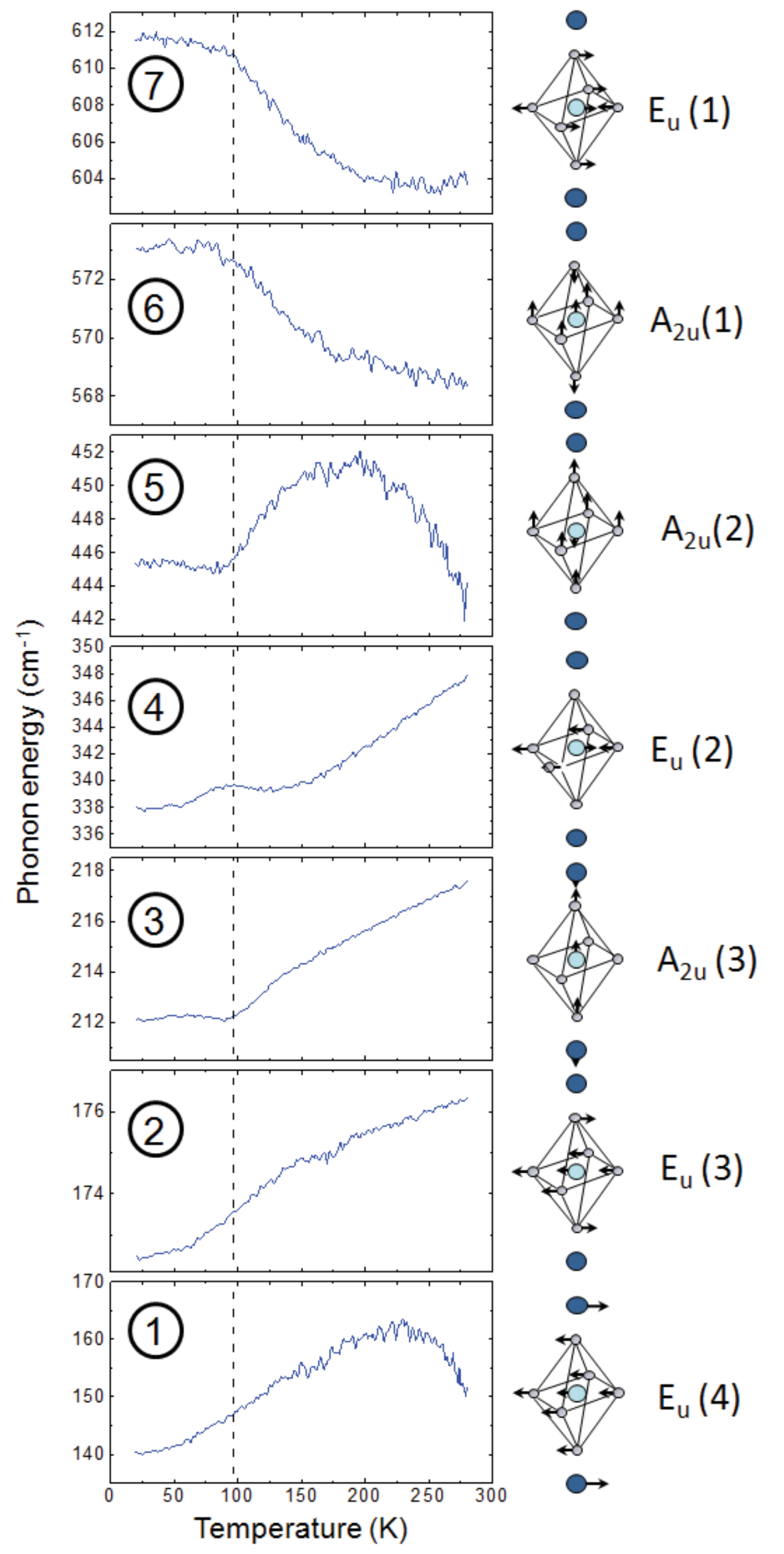}\\
  \caption{(Left panel) Temperature dependence of the energy of phonon modes as numbered in table I. Associated ionic displacement and naming (right panel )}\label{Figphonons}
\end{figure}

Tetragonal Sr$_2$VO$_4$ belongs to the $D_{4h}^{17}$ symmetry group which has 7 infrared, 4 Raman and 1 silent vibration modes. The assignment of the optical phonon modes expected for the $D_{4h}^{17}$ symmetry of  Sr$_2$VO$_4$ has been made from a comparative study of vibration modes in the isostructural compounds Sr$_2$TiO$_4$ \cite{Burns1988} and La$_2$NiO$_4$ \cite{Pintschovius1989}. Optical modes at 142, 174, 213, 340, 445, 516, 574 and 612 cm$^{-1}$ (including some longitudinal components) are indicated by the filled circles in Fig. \ref{fir}. Three additional optical excitations with eigenfrequencies $\omega_1$, $\omega_2$ and $\omega_3$ do not have their counterpart in Sr$_2$TiO$_4$ and La$_2$NiO$_4$ and their origin will be discussed further. The temperature dependence of the phonon modes is shown as colormaps of the optical conductivity in the left panel of Figure ~\ref{Figphonons}.

The parameters used to fit the phonon part of optical data are given in table \ref{param} for three different temperatures.

\begin{table*}
  \centering
  \begin{tabular}{c|cccc|ccc|ccc|c|c}
    \hline\hline
     && \multicolumn{3}{c}{280 K}& \multicolumn{3}{c}{100 K} & \multicolumn{3}{c|}{13 K} &$\Delta\frac{\partial \omega}{\partial T}$&Error on $\Delta\frac{\partial \omega}{\partial T}$\\
   Mode number&$\epsilon_{\infty}$ & \multicolumn{3}{c}{5.83} & \multicolumn{3}{c}{5.99} & \multicolumn{3}{c|}{5.93}&$cm^{-1}.K^{-1}$&$cm^{-1}.K^{-1}$\\
    \hline
&Mode&$\omega_{}$&$\Omega_p$&$\gamma$&$\omega_{}$&$\Omega_p$&$\gamma$&$\omega_{}$&$\Omega_p$&$\gamma$&&\\
1&E$_u$(4)&151&341&116&145&542&76&142&582&75&0.0769&0.1446\\
2&E$_u$(3)&179&298&19&175&418&18&174&427&18&-0.0119&0.0060\\
3&A$_{2u}$(3)&218&382&29&213&403&21&213&409&20&0.0478&0.0071\\
4&E$_u$(2)&349&702&92&341&714&74&340&719&75&-0.0660&0.0268\\
5&A$_{2u}$(2)&443&365&115&445&409&115&445&387&110&0.1449&0.0445\\
5'&E$_u$(2)LO&519&184&38&521&254&42&516&202&32&0.0020&0.0439\\
6&A$_{2u}$(1)&569&310&52&573&334&50&574&341&51&-0.0548&0.0212\\
7&E$_u$(1)&605&188&44&611&167&36&612&180&37&-0.0722&0.0224\\
    \hline\hline
  \end{tabular}
  \caption{Phonon fitting parameters of Sr$_2$VO$_4$ optical data at 280 K, 100 K and 13 K. $\omega$ is the central frequency plotted in Fig.\ref{Figphonons}, $\Omega_p$ is the oscillator strength and $\gamma$ is the scattering rate of the Lorentz oscillator (units are cm$^{-1}$). $\Delta\frac{\partial \omega}{\partial T}$ represents the change in the slope of the phonon energy temperature dependence at the orbital ordering temperature as defined in the text.}\label{param}
\end{table*}

From the Lorentz fit of the different modes, it is possible to follow the evolution of the central frequency upon cooling with a resolution of 1 K (Fig.\ref{Figphonons}). All phonon modes are sensitive to temperature in a different manner depending on the bonds involved and the anisotropy of thermal expansion. We have monitored the variation in the temperature trend of the mode frequency upon crossing the orbital ordering transition. To disentangle the evolution of the phonon frequency due to thermal expansion over the wide temperature range to the effect of orbital rearrangement, we have fitted the temperature dependence of the phonon frequency with a parabolic function above and below the transition temperature $T_1$. We define the change in slopes of the tangents of the 2 fits (above and below transition) at the temperature $T_1$ as:
\begin{equation}\label{slope}
    \Delta\frac{\partial \omega}{\partial T}=\frac{\partial \omega}{\partial T}\Big|_{T_1+\delta}-\frac{\partial \omega}{\partial T}\Big|_{T_1-\delta}
\end{equation}

Three different intervals for each fit are used to calculate error bars. We give in table \ref{param} the value of $\Delta\frac{\partial \omega}{\partial T}$ for all identified phonons. For the soft mode ($1$) and the mode $5'$ (assigned to LO mode), error bars on K are larger than the change in slope. For the remaining modes, we observe a hardening of vibration $3$ and $5$ both involving z-axis displacement ($A_{2u}$ modes) of the vanadium ion opposite to the 4 in plane oxygen ions.

The A$_{2u}$(1) mode (6) also involves displacement of V ion along z-axis but in phase with the 4 in-plane oxygens and thus is only slightly affected by the "in-plane hardening". In-plane modes ($E_{u}$ modes) are softened by the transition. This supports a charge transfer from in-plane to the c-axis orbitals, consequence of a long range orbital ordering.

\subsection{Discussion}

\begin{figure}
  \includegraphics[width=8.5cm]{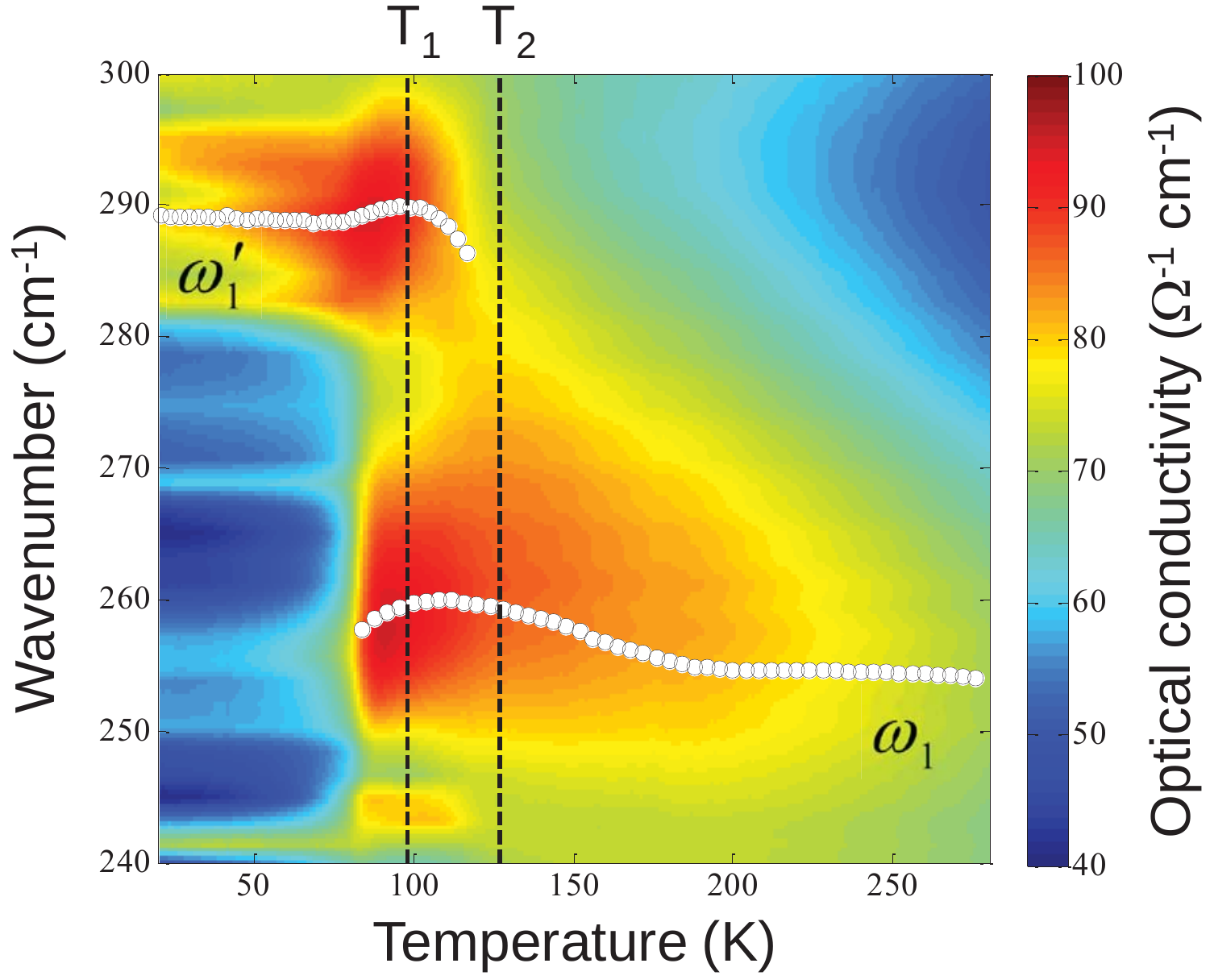}\\
  \caption{Colormap of the experimental optical conductivity in energy range corresponding to the low energy set of transitions. The open symbols represent the temperature dependence of the central frequency of the lorentz oscillator used to fit the reflectivity data. $T_1$ and $T_2$ correspond to the maxima of the two pronounced peaks in the specific heat.}\label{mode1}
\end{figure}

The temperature dependence of the experimental optical conductivity associated with the excitation $\omega_1$ (Fig. \ref{mode1}) is very different in terms of central frequency and spectral weight as compared to the "usual" temperature evolution of the phonons discussed above. Coming from high temperature the eigenfrequency $\omega_1=254$~cm$^{-1}$ ($31.5$~meV) remains almost constant and shifts to a slightly higher value of about 260~cm$^{-1}$ in the phase-coexistence regime. Below about 80~K, $\omega_1$ vanishes. When cooling below $T_2$, an additional mode with a frequency $\omega_1'=290$~cm$^{-1}$ ($36$~meV) emerges and persists down to the lowest temperatures. Note that the simultaneous observation of the two optical excitations in the temperature range 80~K$\leq T\leq T_2$ coincides with the proposed coexistence of two distinct tetragonal phases \cite{Zhou2007}.

Given the known crystal structure of Sr$_2$VO$_4$, we exclude the possibility that $\omega_1$, $\omega_2$ and $\omega_3$ are optical phonons, either from the material itself or from secondary phases that have been the object of a careful study of their own optical properties. The two peaks at $\omega_2$= $810$~cm$^{-1}$ and $\omega_3$= $870$~cm$^{-1}$ are too high in energy to correspond to lattice vibrations.  Moreover, they are close in energy to the twin peaks observed by inelastic neutron scattering at $115$ and $125$~meV \cite{Zhou2010}. With $\Delta E=8$~meV, these excitations are the optical counterparts of the excitations reported by neutron scattering at about $120$~meV with a splitting of about $\Delta E=10$~meV. Energies differ by a factor of about 0.9. This is a natural consequence of the dispersion of the excitations, which for optical spectroscopy are around $k=0$ whereas aforementioned inelastic neutron spectra were integrated over $k$-space.

The frequency of the $\omega_1$ excitation is in good agreement with the calculated frequency ($8.3$ Thz $\simeq 277$ cm$^{-1}$) of the silent $B_{2u}$ mode \cite{Pintschovius1989}. This mode corresponds to vibration along the c-axis of the four oxygen ions in the a-b plane, in phase along the diagonals and with opposite phase along the edges of the a-b oxygen square. The activation of this mode requires a breaking of the crystal symmetry that has so far not been observed in this material.

\begin{figure}
  \includegraphics[width=7.5cm]{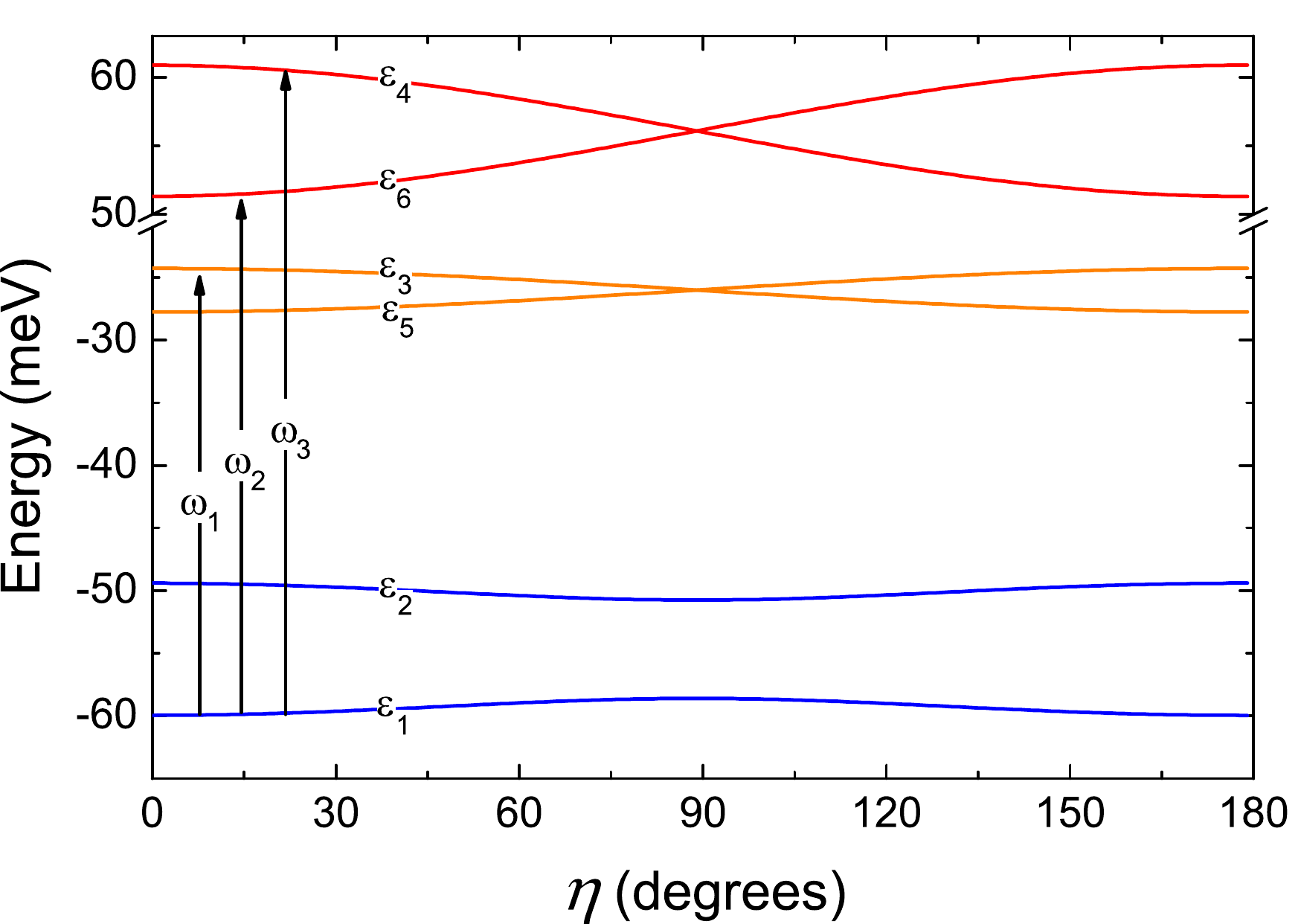}
  \caption{$\eta $-dependence of electron level splitting due to the effects tetragonal crystal field, spin-orbit interaction, antiferromagnetic and ferromagnetic superexchange. Optical transitions are indicated as arrows.}\label{fig:levels}
\end{figure}

Another possibility could be that these three excitations are of electronic origin. $t_{2g}$ splits due to crystal field, spin orbit coupling and exchange fields. We compare our data to a microscopic model for the observed electronic excitation spectrum in Sr$_2$VO$_4$\cite{Eremin2011}. The relevant subset of the $3d$ crystal field excitations is spanned by the $t_{2g}$ levels $d_{xy}$, $d_{xz}$ and $d_{yz}$. The tetragonal crystal field splitting between $d_{xz}$/$d_{xy}$ and $d_{yz}$ is $3D$. The spin-orbit interaction $H_{SO}=\lambda \vec{L}\cdot\vec{S}$ \cite{Abragam1970}, which in d1 systems, tends to align spin and orbital angular momentum opposite to each other. The dominant exchange interaction with the surrounding V$^{4+}$ ions in the $ab$-plane is the anti-ferromagnetic super-exchange, $J_a$, between $d_{xz}$ electrons on nearest neighbor sites along the $x$-axis, and likewise along the $y$-axis. The second largest contribution is the ferromagnetic superexchange, $J_f$, between $d_{xy}$ and $d_{yz}$ along the $x$-axis, and likewise along the $y$-axis. The groundstate is formed by a Kramers doublet having orbital momentum $m_l=\pm 1$  ($\mu_l = 1 \cdot 1\mu_B$) and spin ($\mu_s = 2\cdot \frac{1}{2}\mu_B$) pointing opposite to each other, causing an overall "mute" magnetic moment.  Due to higher order terms in the magnetic exchange interaction the linear combination $u|d_{1,\downarrow}\rangle + v|d_{-1,\uparrow}\rangle$ with $u=cos(\eta/2)$ and $v=sin(\eta/2)$ has a weak dependence on $\eta$ as shown in Fig.~\ref{fig:levels}. The magnetically most stable state corresponds to $\eta=0$.

The energy levels can be exactly calculated from the model and they are drawn as a function of the order parameter $\eta $ in Fig.~\ref{fig:levels}. The arrows represent the possible electronic transitions that could match with the $\omega_1$, $\omega_2$ and $\omega_3$ transitions observed in optical spectra.
Using the above level scheme for $\eta =0$ (Fig. \ref{fig:levels})  we now obtain a consistent description of the observed optical excitation $\omega_1'=290$~$cm^{-1}$ ($\simeq 36$~meV) and a double peak structure (Fig. \ref{optics}b) with central
frequencies of $\omega_2=810$~$cm^{-1}$ ($100$~meV) and $\omega_3=870$~$cm^{-1}$ ($108$~meV) at $20$~K.

Using a value $\lambda \simeq -30$~meV equal to the free-ion value \cite{Abragam1970}, $J_a\simeq15$~meV, $D=-33$~meV and $J_f\simeq -9$~meV in the model described above, we get a good estimate for the excitation $\omega_{1}$. The excitation pair $\omega_{2,3}$ between the lowest and the highest doublet is mainly determined by the crystal-field parameter and the difference in energy $\omega_{3}-\omega_{2}$ agrees with the splitting of 10~meV in the neutron scattering spectra. The fact that this splitting reportedly persists up to 400~K indicates that short-range ordering is present in the system far
above room temperature and the level scheme should be valid even
above the phase transition. We estimate that the
reduction of $D$ due to the reported reduced $c/a$-ratio in the
high-temperature tetragonal structure \cite{Zhou2007} is less than
one percent. Therefore, we ascribe the energy difference of 4~meV of
the optical excitation between the low- and high-temperature
tetragonal structures and the decreasing splitting of the highest
lying levels to a somewhat smaller contribution of the spin-orbit
interactions in the short-range ordered regime due to reduced
intersite spin-spin correlations.\\

Having consistently described the observed optical excitations using this model, the remaining question is why
these excitations are optically active. Given the symmetries of the Sr$_2$VO$_4$ as they have been described in the literature, and as we have assumed them to be in our discussion, the ground state and the excited states within the $t_{2g}^1$ manifold all have even parity. Consequently under this assumption none of the three excitations discussed above are expected to be electric-dipole active. Instead, since they are of the magnetic-dipole variety, we expect a relatively small oscillator strength \cite{Kant2008,TanabeSuganoBook}. For an isolated ion with $\eta =0$, the selection rule $\Delta J_z=\pm 1$ applies to the optical transitions at $\omega_1$ and $\omega_3$. The excitation at $\omega_2$ requires $\Delta J_z=2$, which corresponds to a quadrupole transition. The admixture of $|d_{1,\downarrow}\rangle$ and $|d_{-1,\uparrow}\rangle$ in the ground state gives a dipole like $\Delta J_z=\pm 1$ for the excitation $\omega_2$. An alternative possibility is a mechanism described by \textcite{Tanabe1965}, whereby a finite electric-dipole matrix element is induced by the breaking of inversion symmetry of a {\em pair} of neighboring spins, as observed in FeF$_2$\cite{Halley1965}.\\

\section{Conclusions}
We have measured the infrared and visible optical spectrum of tetragonal Sr$_2$VO$_4$ at low temperature. In addition to well identified phonon bands, an additional peak was observed, the origin of which could be :
\begin{itemize}
\item{A non optically active phonon that is made IR-active due to beaking of crystal field symmetry}
\item{The electronic excitation spectrum of the V$^{4+}$ ions that supports a scenario of a novel ordered state in terms of an alternating spin-orbital order in Sr$_2$VO$_4$.}
\end{itemize}
In this last scenario, the magnetic moments are muted by spin-orbit interaction. At low temperature these mute moments are anti-ferromagnetically ordered. At elevated temperatures the long range order is lost, resulting in a high temperature phase which is again tetragonal. Since the ordering involves not only spin, but also orients the angular moments of the ions, the resulting magnetostriction should be particularly strong. This is probably the reason for the change of lattice constant when the system orders and for the coexistence of two thermodynamically distinct phases in a certain temperature range.
\acknowledgements
We gratefully acknowledge D. I. Khomskii and G. Jackeli for helpful and stimulating discussions. This work is supported by the SNSF through Grant No. 200020-130052 and the National Center of Competence in Research (NCCR) "Materials with Novel Electronic Properties-MaNEP". We acknowledge partial support by the DFG via the Collaborative Research Center TRR 80. MVE is partially supported by the Ministry of Education of the Russian Federation via Grant No. 1.83.11.

\end{document}